\documentclass[twocolumn]{aastex701}
\usepackage{graphicx}
\usepackage{subcaption}
\usepackage{threeparttable}

\shorttitle{EP241021a HST}
\shortauthors{Sun et al.}

\begin{document}

\title{EP241021a as An Off-nuclear Month-duration Transient in a Dwarf Starburst Galaxy}

\author[orcid=0000-0002-7223-5840,sname='Sun']{Luming Sun}
\affiliation{Department of Physics, Anhui Normal University, Wuhu, 241002, China}
\affiliation{Center for Astrophysics and Astronomical Technology, Anhui Normal University, Wuhu, 241002, China}
\email[show]{sunluming@ahnu.edu.cn}

\author[orcid=0000-0002-7020-4290,sname='Shu']{Xinwen Shu}
\affiliation{Department of Physics, Anhui Normal University, Wuhu, 241002, China}
\affiliation{Center for Astrophysics and Astronomical Technology, Anhui Normal University, Wuhu, 241002, China}
\email[show]{xwshu@ahnu.edu.cn}

\author[orcid=0000-0003-3824-9496,sname='Zhu']{Jiazheng Zhu}
\affiliation{Department of Astronomy, University of Science and Technology of China, Hefei, Anhui 230026, China}
\affiliation{School of Astronomy and Space Sciences, University of Science and Technology of China, Hefei 230026, People’s Republic of China}
\email{jiazheng@mail.ustc.edu.cn}

\author[orcid=0000-0002-7152-3621,sname='Jiang']{Ning Jiang}
\affiliation{Department of Astronomy, University of Science and Technology of China, Hefei, Anhui 230026, China}
\affiliation{School of Astronomy and Space Sciences, University of Science and Technology of China, Hefei 230026, People’s Republic of China}
\email{jnac@ustc.edu.cn}

\author[orcid=0000-0002-9725-2524,sname='Zhang']{Bing Zhang}
\affiliation{The Hong Kong Institute for Astronomy and Astrophysics, the University of Hong Kong, Pokfulam Road, Hong Kong, China}
\affiliation{Department of Physics, The University of Hong Kong, Pokfulam Road, Hong Kong, China}
\email{bzhang1@hku.hk}

\author[orcid=0000-0002-3525-791X,sname='Chen']{Jin-Hong Chen}
\affiliation{Department of Physics, The University of Hong Kong, Pokfulam Road, Hong Kong, China}
\affiliation{The Hong Kong Institute for Astronomy and Astrophysics, University of Hong Kong, Hong Kong, China}
\affiliation{Shenzhen Institute of Research and Innovation, The University of Hong Kong, Shenzhen 518057, P. R. China}
\email{chenjh2@hku.hk}

\author[orcid=0000-0002-9589-5235,sname='Dai']{Lixin Dai}
\affiliation{Department of Physics, The University of Hong Kong, Pokfulam Road, Hong Kong, China}
\email{lixindai@hku.hk}

\author[orcid=0009-0001-8733-2088,sname='Ding']{Hucheng~Ding}
\affiliation{Department of Physics, Anhui Normal University, Wuhu, 241002, China}
\affiliation{Center for Astrophysics and Astronomical Technology, Anhui Normal University, Wuhu, 241002, China}
\email{dinghc@ahnu.edu.cn}

\author[sname='Jin']{Chichuan Jin}
\affiliation{National Astronomical Observatories, Chinese Academy of Sciences, Beijing 100101, China}
\affiliation{School of Astronomy and Space Science, University of Chinese Academy of Sciences, Beijing 100049, China}
\affiliation{Institute for Frontier in Astronomy and Astrophysics, Beijing Normal University, Beijing 102206, China}
\email{ccjin@nao.cas.cn}

\author[orcid=0000-0002-1568-7461
,sname='Lu']{Wenbin Lu}
\affiliation{Department of Astronomy and Theoretical Astrophysics Center, University of California, Berkeley, CA 94720, USA}
\email{wenbinlu@berkeley.edu}

\author[orcid=0000-0001-5012-2362,sname='Shen']{Rong-Feng Shen}
\affiliation{School of Physics and Astronomy, Sun Yat-Sen University, Zhuhai 519082, China}
\email{shenrf3@mail.sysu.edu.cn}

\author[orcid=0000-0001-7943-4685,sname='Xu']{Fan Xu}
\affiliation{Institute of Space Weather, School of Atmospheric Physics, Nanjing University of Information Science and Technology, Nanjing 210044, China}
\email{fanxu@nuist.edu.cn}

\author[sname='Yang']{Lei Yang}
\affiliation{Department of Physics, Anhui Normal University, Wuhu, 241002, China}
\affiliation{Center for Astrophysics and Astronomical Technology, Anhui Normal University, Wuhu, 241002, China}
\email{ahnuyi@ahnu.edu.cn}

\author[orcid=0000-0002-1067-1911,sname='Yu']{Yunwei Yu}
\affiliation{Institute of Astrophysics, Central China Normal University, Wuhan 430079, China}
\email{yuyw@ccnu.edu.cn}

\author[sname='Yuan']{Weimin Yuan}
\affiliation{National Astronomical Observatories, Chinese Academy of Sciences, Beijing 100101, China}
\email{wmy@nao.cas.cn}

\author[orcid=0000-0002-4562-7179,sname='Li']{Dongyue Li}
\affiliation{National Astronomical Observatories, Chinese Academy of Sciences, Beijing 100101, China}
\email{dyli@nao.cas.cn}

\author[orcid=0000-0003-3257-9435,sname='Xu']{Dong Xu}
\affiliation{National Astronomical Observatories, Chinese Academy of Sciences, Beijing 100101, China}
\affiliation{Altay Astronomical Observatory, Altay, Xinjiang 836500, China}
\email{dxu@nao.cas.cn}

\author[orcid=0000-0002-1517-6792,sname='Wang']{Tinggui Wang}
\affiliation{Department of Astronomy, University of Science and Technology of China, Hefei, Anhui 230026, China}
\affiliation{School of Astronomy and Space Sciences, University of Science and Technology of China, Hefei 230026, People’s Republic of China}
\email{twang@ustc.edu.cn}

\author[sname='Yang']{Haonan Yang}
\affiliation{National Astronomical Observatories, Chinese Academy of Sciences, Beijing 100101, China}
\email{hnyang@nao.cas.cn}

\author[sname='Zhang']{Wenda Zhang}
\affiliation{National Astronomical Observatories, Chinese Academy of Sciences, Beijing 100101, China}
\email{wdzhang@nao.cas.cn}

\begin{abstract}

EP241021a is a peculiar fast X-ray transient exhibiting a long X-ray plateau and an optical rebrightening, both challenging the traditional jet afterglow model.
Whether its nature is a stellar explosion or a tidal disruption event (TDE) remains debated.
Here we present late-time HST/WFC3 and Gemini-N/GMOS observations of its host galaxy at $z=0.748$.
After $\sim300$ days post-trigger, the transient emission had faded below detection, revealing a dwarf disk galaxy with an effective radius $R_e=1.7$ kpc, a stellar mass $M_\star\sim3\times10^8M_\odot$, and a high star formation rate of $\sim5\ M_\odot\ {\rm yr}^{-1}$.
Combined with previous very long baseline interferometry measurements, the transient can be located in the galactic disk with a projected physical offset of $0.95\pm0.13$ kpc from the galaxy center.
Host-subtracted optical light curves show that the rebrightening ceased at $\sim47$ days, followed by a sudden drop with a power-law index steeper than $-3.3$, coincident with a similar drop in the X-ray light curve.
This synchronous drop in both bands is consistent with jet shut-off, and the required month-long jet activity suggests a persistently accreting compact object fed by a TDE.
Among the TDE scenarios, the off-nuclear location and the starburst host galaxy slightly favor the post-supernova micro-TDE scenario in a binary system, in which a natal kick drives the newborn compact object into a tidal disruption of its companion.
Nevertheless, an off-nuclear intermediate-mass black hole TDE or a stellar-mass black hole TDE induced by dynamical scattering in a dense stellar cluster cannot yet be excluded.

\end{abstract}


\section{Introduction} \label{sec:1}

Fast X-ray Transients (FXTs) are a class of extragalactic astrophysical phenomena characterized by intense, brief bursts of soft X-ray emission, with durations ranging from tens to thousands of seconds and peak isotropic luminosities spanning a broad range from $10^{42}$ to $10^{50}$ erg s$^{-1}$ \citep{Heise2003,Soderberg2008,Jonker2013,Glennie2015,Bauer2017,Xue2019,Alp2020,Lin2022,Quirola-Vasquez2022,Quirola-Vasquez2023}.
Unlike traditional Gamma-Ray Bursts (GRBs), most FXTs are not detected in the gamma-ray band, constituting a subclass of ``gamma-ray dark'' transients \citep{ZhangYQ2025}.

Although dozens of FXTs have been identified, their physical origins remain unclear and several competing theoretical models have been proposed.
First, the high-redshift GRB model argues that cosmological redshift shifts the peak energy of GRB spectra into the soft X-ray band, causing them to appear as gamma-ray-dark FXTs \citep{Heise2003}.
Second, the massive star collapse model suggests that FXTs can originate from the shock breakout of core-collapse supernovae \citep{Soderberg2008,Alp2020}.
Third, the compact object merger model proposes that mergers of binary neutron stars form millisecond magnetars, which trigger FXTs when they spin down \citep{Xue2019,Lin2022}.
Fourth, the tidal disruption event (TDE) model posits that the tidal disruption of a white dwarf star by an intermediate-mass black hole (IMBH) triggers super-Eddington accretion, which in turn drives a relativistic jet and produces an X-ray flare \citep{Jonker2013,PengZK2019}.

The Einstein Probe (EP; \citealt{YuanWM2025}) has greatly expanded the multi-wavelength observational sample of FXTs.
EP has revealed several landmark events, such as EP240315a, EP240414a, and EP250108a \citep{LiuY2025,SunH2025,Rastinejad2025}.
Furthermore, EP has discovered a new class of FXTs characterized by X-ray plateaus lasting several days or longer.
The first such case is EP240408a, which exhibits a plateau lasting about 4 days in its X-ray light curve, followed by a sudden drop to undetectable levels \citep{ZhangWD2025,OConnor2025}. 
However, follow-up observations in optical, near-infrared, and radio bands did not detect any 
counterparts to EP240408a, making its nature largely unknown.

EP241021a at $z=0.748$ is a unique case with an even longer X-ray plateau and rich multi-wavelength emission, enabling a detailed study of its complex evolution \citep{Shu2025,Busmann2025,Gianfagna2025,Yadav2025,Quirola-Vasquez2026}.
It was triggered by EP/WXT on October 21, 2024, producing an X-ray flare lasting approximately 100 s with a peak luminosity of $\sim10^{48}$ erg s$^{-1}$, but no gamma-ray counterpart was detected by monitors such as Fermi-GBM and Konus-Wind.
The high-energy spectral energy distribution suggests a peak energy $E_p\lesssim100$ keV and an isotropic-equivalent energy $E_{\rm iso}\lesssim2\times10^{51}$ erg.
Subsequent follow-up observations extending to beyond 100 days have revealed a series of anomalous features.
The X-ray emission lasts for more than one month (one of the longest FXT afterglows), and exhibits a complex evolution with an initial decay, a plateau, a rebrightening, and a late-time steep decline.
The optical light curve shows multiple rebrightening episodes, peaking at an absolute magnitude of $M_r\approx-22$ mag, and exhibiting persistently red colors ($r-J\approx0.8$).
The radio emission exhibits rapid spectral evolution and a double-peaked spectral structure at $\sim70$ days, suggesting the presence of multiple emission components.
Photometry and spectroscopy of the host galaxy reveal a low-mass, young galaxy with a high star formation rate.
These observational facts pose serious challenges to any single-component models, and several independent works have proposed different explanations, including multiple jet ejections \citep{Shu2025}, a refreshed shock \citep{Busmann2025}, an off-axis structured jet with a two-layer cocoon \citep{Gianfagna2025}, a magnetar-powered thermal ejecta \citep{Wu2025}, and a forward shock plus supernova model \citep{Quirola-Vasquez2026}.
Therefore, EP241021a has become a key testbed for understanding the diversity and the unified physical mechanisms of FXTs.

In this paper, we present new Hubble Space Telescope (HST) and ground-based observations of the host galaxy of EP241021a, which clearly reveal that the transient locates at an off-nuclear region.
The paper is organized as follows.
Section 2 describes the observations and data reduction.
Section 3 presents the imaging and spectral energy distribution analysis of the host galaxy, as well as the host-subtracted optical light curve of the transient.
In Section 4 we discuss the nature of EP241021a by combining the new results obtained in this work with previous observations.
Our main conclusions are summarized in Section 5.
Throughout this paper, we adopt a cosmology of $\Omega_M=0.3$, $\Omega_\Lambda=0.7$, and $H_0=70$ km s$^{-1}$ Mpc$^{-1}$. Thus, the inferred luminosity distance is $D_L=4.620$ Gpc and $1\arcsec$ corresponds to a physical size of 7.330 kpc.
All phases are calculated relative to the WXT discovery at ${\rm MJD}=60604.2$.
All magnitude measurements were reported in the AB system without correction for galactic extinction, and we will correct them using $A_V=0.139$ for subsequent analysis.

\section{Observation and data reduction} \label{sec:2}

To monitor the late-time evolution of EP241021a, we observed it with the Gemini Multi-Object Spectrographs (GMOS) mounted on the Gemini-North telescope on September 3, 2025, at a phase of 316 days.
We chose the $r$ and $z$ filters, taking four dithered frames in each band, each with an exposure time of 900 s.
We performed standard bias and flat corrections, and stacked the frames after astrometric calibrations against the Panoramic Survey Telescope \& Rapid Response System (PS1) catalog\footnote{https://catalogs.mast.stsci.edu/panstarrs/} using \texttt{Sextractor} and \texttt{Scamp} \citep{Bertin1996,Bertin2006}.
EP241021a was not resolved in either band, with a spatial resolution of $\sim0.7\arcsec$.
After flux calibration using the PS1 catalog, we measured $r=24.86\pm0.05$ and $z=23.78\pm0.11$ using automatic aperture photometry.
The $r$-band magnitude is consistent with that observed at a phase of 94 days ($R=24.8\pm0.1$; \citealt{Busmann2025}), implying that the transient had faded and the flux was dominated by the host galaxy.

To further constrain the morphology and physical parameters of the host galaxy, and to accurately measure the position of the galaxy center and compare it with the transient position, we triggered an HST Director's Discretionary Time (DDT) observation of EP241021a with the Wide Field Camera 3 (WFC3; \citealt{HST_WFC3}).
We used four filters: F606W, F814W, F110W, and F160W.
The observations in the F606W and F814W filters were carried out on November 3, 2025, while those in the F110W and F160W filters were obtained on June 21, 2026.
The exposure times for the four filters were 2080, 840, 256, and 256 seconds, respectively.
We downloaded images that had been processed by the HST/WFC3 pipeline\footnote{https://hst-docs.stsci.edu/wfc3dhb} from the Barbara A. Mikulski Archive for Space Telescopes\footnote{https://archive.stsci.edu/} (MAST).
The pipeline performed bias, dark and flat corrections, corrected distortions, combined dithered exposures, and applied flux calibration.

For the F606W and F814W observations, we re-performed the astrometric calibration to improve the positional accuracy in comparison with the pipeline calibration.
During the observations, we included four astrometric standard stars in the field together with EP241021a.
We selected them from the Gaia DR3 catalogue \citep{GaiaDR3} with $R$-band magnitudes ranging from 18.4 to 20.0, to ensure that they would not saturate during a single exposure.
We assumed a linear transformation converting the target position measured on the image into celestial coordinates, which includes three free parameters: deviations in central RA and Dec relative to pipeline calibration, and a deviation of the rotation angle.
For the four standard stars, we calculated the offsets between the coordinates converted with the transformation and those predicted by the Gaia catalogue.
We determined the three parameters by minimizing the sum of the squared offsets.
We estimated the systematic error using a leave-one-out method.
For each standard star, we determined the transformation parameters using the remaining three stars and converted its position.
We then calculated the offset between this converted coordinate and that predicted by the Gaia catalogue.
The root mean square (rms) of the four offsets was adopted as the systematic error.
The systematic errors are 19 and 28 mas for the F606W and F814W bands, respectively.

\section{Data analysis and results} \label{sec:3}

\subsection{HST image} \label{sec:3.1}

\begin{table*}
\centering
\caption{Image fitting results}
\begin{tabular}{cccccccccc}
\hline
\hline
           & RA center & DEC center & $\sigma_{\rm stat}$ & $\sigma_{\rm sys}$ & $\sigma_{\rm tot}$ & $R_e$ & $b/a$ & $\theta_{\rm maj}$ &Magnitude \\
           & (1)       & (2)        & (3)                 & (4)                & (5)                & (6)   & (7)   & (8)   & (9) \\
\hline
F606W         & 01:55:23.4396 & +05:56:17.863 & 4.7  & 19 & 20 & 232 & 0.297 & 48.5 & $25.101\pm0.037$ \\
F814W         & 01:55:23.4394 & +05:56:17.881 & 10.6 & 28 & 30 & 231 & 0.353 & 47.0 & $24.452\pm0.082$ \\
Weighted mean & 01:55:23.4396 & +05:56:17.869 &      &    & 17 &     &      &  & \\
\hline
\end{tabular}
\begin{tablenotes}
    \item (1),(2): The central position; (3),(4),(5): The statistical, systematic and the total error of the central position in unit of mas; (6): The effective radius in unit of mas; (7): The ratio of the semiminor and semimajor axes; (8): The inclination angle of the major axis (degree, north to east); (9): The AB magnitude of the galaxy.
\end{tablenotes}
\label{tab:hst_param}
\end{table*}

\begin{figure}
\centering
  \includegraphics[scale=0.83]{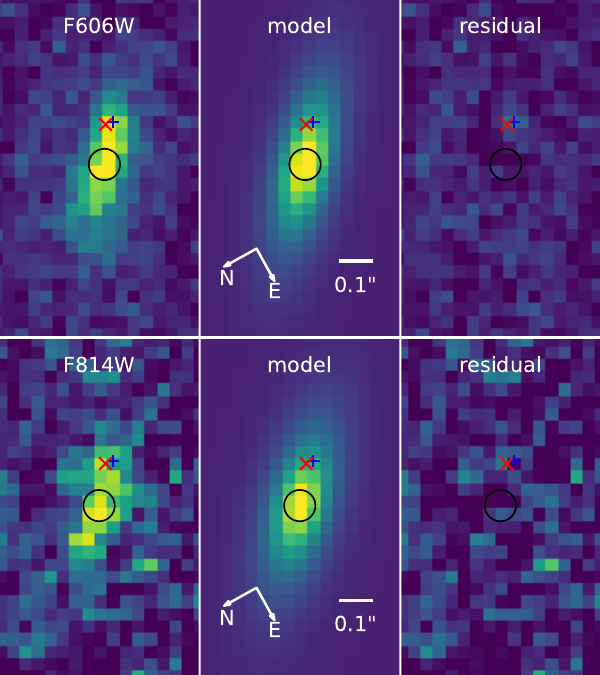}
  \caption{
  Image data (left), best-fitting exponential models (middle) and residuals (right) in the F606W (top) and F814W (bottom) bands.
  We labeled the galaxy center (black open circle), along with the transient position measured from VLBA (red cross) and e-MERLIN (blue plus).
  The VLBA position is RA = $01^{\rm h} 55^{\rm m} 23^{\rm s}.4323$, DEC = $+05^\circ 56^\prime 17.798^{\prime\prime}$ with a $1\sigma$ uncertainty of (4.9, 1.5) mas in RA and DEC, respectively.
  The e-MERLIN position is RA = $01^{\rm h} 55^{\rm m} 23^{\rm s}.4325$, DEC = $+05^\circ 56^\prime 17.774^{\prime\prime}$ with an uncertainty of (9, 3) mas.
  }
\label{fig:HSTimage}
\end{figure}

We measured the morphology and center position of EP241021a's host galaxy using the F606W and F814W images, which have a high spatial resolution of $\sim70$ mas\footnote{https://hst-docs.stsci.edu/wfc3ihb/} and improved astrometric calibration.
Figure~\ref{fig:HSTimage} presents the image of the host galaxy.
It appears to be a disk galaxy viewed edge-on.
We then fitted the two-dimensional brightness profile using an exponential model with elliptical isophotes.
As shown in Figure~\ref{fig:HSTimage}, this model provides good fits to the data in both bands, and the residuals are consistent with the background noise.
We summarize the parameters in Table~\ref{tab:hst_param}.
The fitting confirms that the host is a disk galaxy.
Its effective radius $R_e$ is $0.23\arcsec$ ($\sim1.7$ kpc), consistent with its being unresolved in the Gemini observation with a spatial resolution of $0.7\arcsec$.

In both bands, the positional uncertainties of the galaxy center are dominated by systematic errors.
Using a weighted average, we determined the galaxy center to be at RA = $01^{\rm h} 55^{\rm m} 23^{\rm s}.4396$, DEC = $+05^\circ 56^\prime 17.869^{\prime\prime}$ (J2000) with a total uncertainty of 17 mas at $1\sigma$ confidence.
As shown in Figure~\ref{fig:HSTimage}, the galaxy center deviates significantly from the transient position measured by the Very Long Baseline Array (VLBA; \citealt{Shu2025}) and e-MERLIN (\citealt{Yadav2025}).
Using the more accurate VLBA measurement, we obtained an offset of $129\pm18$ mas, corresponding to a projected physical offset of $0.95\pm0.13$ kpc.
The transient position is consistent with the galaxy's major axis within the uncertainties, suggesting that it may occur in the galaxy's disk.

To place a robust upper limit on the optical emission from the transient at the HST epoch, we fit the F606W and F814W images with an exponential disk model plus an additional point source component at the fixed VLBA position of the transient.
We obtained $5\sigma$ magnitude limits of ${\rm F606W}>28.45$ and ${\rm F814W}>27.03$, confirming that the transient emission had faded below detection at 379 days post-trigger.

\subsection{Spectral energy distribution of the host galaxy} \label{sec:3.2}

\begin{figure}
\centering
  \includegraphics[scale=0.83]{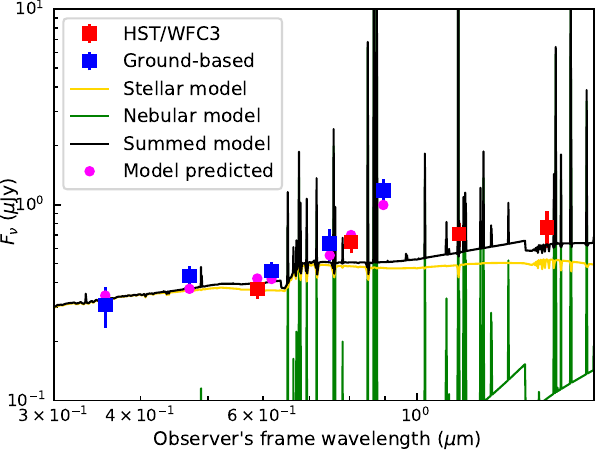}
  \caption{
  The SED of the host galaxy and the best-fitting CIGALE model.
  Red and blue squares indicate data from HST and ground-based observations.
  The stellar and nebular components, and the summed model, are shown as yellow, green, and black lines, respectively.
  The model predicted photometries are shown as magenta points for comparison.
  }
\label{fig:sedfit}
\end{figure}

To estimate the properties of the host galaxy, we constructed its spectral energy distribution (SED) using photometric measurements from HST and ground-based observations.
The best-fit model magnitudes in the F606W and F814W bands are $25.10\pm0.04$ and $24.45\pm0.08$, respectively.
We also fitted the F110W and F160W images with the same method, yielding model magnitudes of $24.32\pm0.10$ and $24.21\pm0.21$, respectively.
We adopted $u=25.40\pm0.23$, $g=24.98\pm0.09$, and $i=24.48\pm0.17$ from \cite{Quirola-Vasquez2026}, and $r=24.86\pm0.05$ and $z=23.78\pm0.11$ from our Gemini-N observations, which have improved photometric accuracy.
The resulting SED is shown in Figure~\ref{fig:sedfit}.

\begin{figure*}
\centering
  \includegraphics[width=1.0\textwidth]{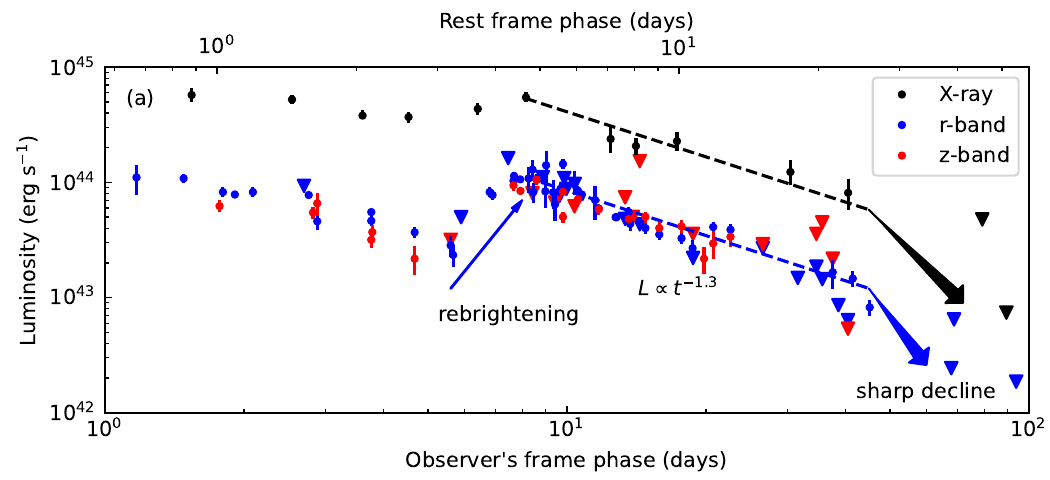}
  \includegraphics[width=0.4\textwidth]{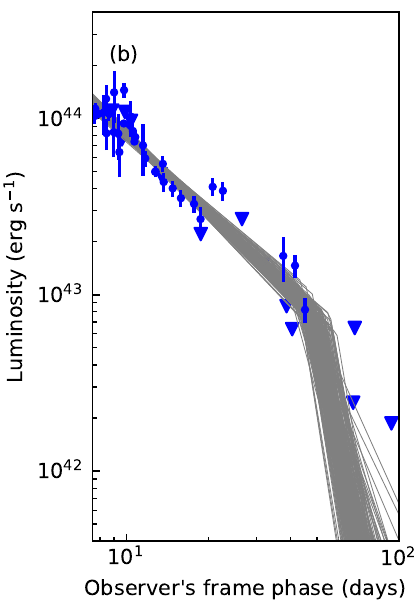}
  \includegraphics[width=0.59\textwidth]{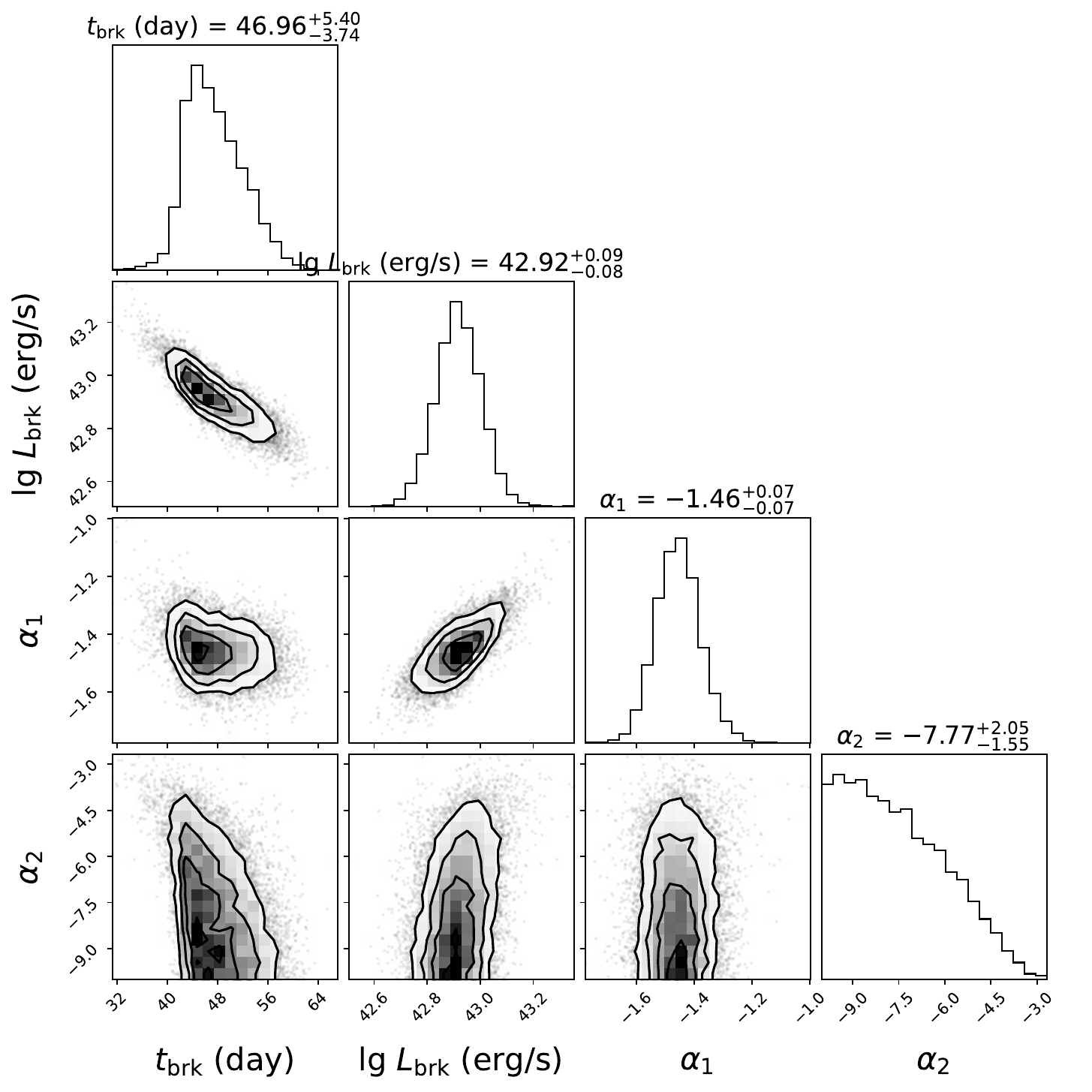}
  \caption{
  \textbf{(a)}: The host-subtracted optical light curves of the transient EP241021a in the $r$ and $z$ bands, along with the X-ray light curves during and after the plateau phase.
  The data were taken from \cite{Shu2025} and \cite{Busmann2025}.
  The rebrightening, the subsequent power-law decay, and the final sharp decline are labeled in both bands.
  \textbf{(b)}: The broken power-law models that fit the $r$-band data after a phase of 7.5 days.
  We present 200 randomly selected MCMC realizations for clarity.
  \textbf{(c)}: The posterior distributions for each model parameter.
  }
\label{fig:subhost}
\end{figure*}

We fit the SED using the Code Investigating GALaxy Emission (CIGALE; \citealt{Boquien2019}).
We assumed a delayed exponential decay for the star formation history, and adopted the initial mass function of \cite{Salpeter1955} and the single stellar population templates of \cite{Bruzual2003}, with the stellar metallicity ($Z_\star$) as a free parameter.
We adopted dust attenuation and emission models from \cite{Calzetti2000} and \cite{Dale2014}.
We also included nebular emission, leaving the gas-phase metallicity ($Z_{\rm gas}$) free while keeping the other nebular parameters at their default values.
Fixing $Z_{\rm gas}$ at the solar value (0.02) led to a [OIII] luminosity lower than the spectroscopic value of \cite{Shu2025} and a discrepancy in the $z$ band.

The model provided a good fit to the SED data with a reduced $\chi^2$ of 0.7 after adding a systematic error of 10\%.
The model [OIII] luminosity of $\sim5\times10^{41}$ erg~s$^{-1}$ is consistent with the spectroscopic value of $3.6\times10^{41}$ erg~s$^{-1}$, given the scatter among the model SEDs and the spectral calibration uncertainties.
The best-fit model yields a stellar mass $M_\star=2.7\pm1.2\times10^8\ M_\odot$, a star formation rate ${\rm SFR}=4.5\pm2.4\ M_\odot\ {\rm yr}^{-1}$, an age $t=42\pm22$ Myr, a delay time $\tau=54\pm31$ Myr, and a dust attenuation $A_V=0.69\pm0.16$ for the stellar continuum.
The stellar mass and star formation rate are consistent with the measurements of \cite{Shu2025} and \cite{Quirola-Vasquez2026} within the uncertainties, classifying the host galaxy as a dwarf starburst galaxy.
The model requires sub-solar metallicities, with $3\sigma$ upper limits of $Z_{\rm star}<0.02$ and $Z_{\rm gas}<0.01$, as expected from the mass-metallicity relation of star-forming galaxies \citep{Kewley2008}.

\subsection{Host-subtracted optical light curve of the transient}

Previous studies have reported a prominent optical rebrightening at a phase of $\sim7$ days \citep{Busmann2025}.
However, the subsequent evolution of the optical emission remained uncertain, as the host contribution was not subtracted.
With reliable photometric data for the host galaxy, we re-analyze the host-subtracted optical light curve.
We converted the early-time $r-$ and $z-$band magnitudes reported in the literature to fluxes and subtracted the host fluxes derived from the Gemini-N photometry at a phase of 316 days.
All photometries were obtained with ground-based telescopes with PSF FWHM of $\gtrsim0.5\arcsec$, much larger than the effective radius of the host galaxy ($0.23\arcsec$).
The host is therefore unresolved in all these images, which justifies subtracting the host flux directly from the total flux.
The host-subtracted light curves of EP241021a is presented in Figure~\ref{fig:subhost}(a).

No significant optical emission from the transient was detected at phases later than $\sim45$ days.
We obtained a deep $3\sigma$ upper limit on the optical emission at 68 days, with $\nu L_\nu(r)<2.4\times10^{42}$ ${\rm erg\ s^{-1}}$, indicating that the optical flux had already declined sharply by that epoch.

To quantify this decline, we fitted the r-band light curve after the rebrightening (phase $>7.5$ days) with a broken power-law model, following the method of \cite{Shu2025} for the X-ray light curve.
The model parameters include the break time $t_{\rm brk}$, the luminosity at the break $L_{\rm brk}$, and the power-law indices $\alpha_1$ and $\alpha_2$ before and after the break.
We adopted a logarithmically uniform prior for $L_{\rm brk}$ and uniform priors for the other parameters, with $\alpha_2>-10$ as a prior constraint.
With the Markov Chain Monte Carlo (MCMC) code \texttt{emcee}, we obtained $t_{\rm brk}=47.0^{+5.4}_{-3.7}$ days, ${\rm lg}\ L_{\rm brk}=42.92\pm0.08$ ${\rm erg\ s^{-1}}$, $\alpha_1=-1.46\pm0.07$, and $\alpha_2=-7.8^{+2.1}_{-1.6}$, as can be seen from Figure~\ref{fig:subhost}(b) and \ref{fig:subhost}(c).
The value of $\alpha_2$ is influenced by the prior, but a $3\sigma$ upper limit gives $\alpha_2<-3.3$.
Thus, the rebrightening lasted until $\sim47$ days, after which the optical flux declined sharply.

\section{Discussion} \label{sec:4}

\subsection{Optical and X-ray drop: jet shut-off?}

In the host-subtracted optical light curve, the flux exhibits a sudden drop at a phase of $26.9^{+3.1}_{-2.1}$ days in the rest frame, with the decline steepening from $\alpha_1\sim-1.5$ to $\alpha_2<-3.3$.
As first noticed by \cite{Shu2025}, the X-ray light curve also shows a sharp drop at a rest-frame break time of $33\pm10$ days (see Fig.~\ref{fig:subhost}), consistent with the optical drop.
Such a break and steep decline are inconsistent with the standard afterglow from a supernova-associated jet, which typically decays with a power-law index shallower than $-3$ \citep[e.g.,][]{Kumar2000}, and require an additional explanation.

Such a sudden drop is reminiscent of the jet shut-off phenomenon observed in jetted TDEs.
Among the four known supermassive black hole TDEs (SMBH) with relativistic jet, three---Swift J1644+57, Swift J2058+05, and AT2022cmc---exhibit a sharp break in their X-ray light curves at rest-frame times of $\sim100-400$ days with X-ray luminosities of $\sim10^{44}-10^{45}$ erg~s$^{-1}$, followed by a rapid decline by 1--2 orders of magnitudes \citep{Zauderer2013,Pasham2015,Eftekhari2024}.
For the fourth source, Swift J1112-82, deep late-time X-ray data are lacking to confirm a similar shut-off.
This rapid decline is inconsistent with a forward-shock origin and instead points to the cessation of jet activity when the accretion rate drops below Eddington.
EP241021a shows a similar sudden drop in both the optical and X-ray light curves, supporting a similar jet shut-off origin.
The earlier rest-frame break time ($\sim30$ days) and lower break luminosity (a few $10^{43}$ erg~s$^{-1}$ and $\sim10^{43}$ erg~s$^{-1}$ in the X-ray and optical bands, respectively) may reflect a difference in the central engine \citep{Shu2025}.

Alternatively, the optical break may not be intrinsic to the jet but instead arises from an additional thermal component.
A possible slight optical rebrightening at $\sim20$ days has been suggested, after which the optical SED can be described by SN thermal emission \citep{Wu2025,Quirola-Vasquez2026}.
This could be attributed to an SN with a luminosity and evolution time scale similar to SN 1998bw (\citealt{Quirola-Vasquez2026}, their Fig. 8).
In this case, the underlying jet component could decline with a steeper power-law than previously measured ($\alpha_1\sim-1.5$), and hence extend smoothly to the late-time low luminosity level.
The apparent optical break would then be an artifact of the superposition of the jet and thermal components.
However, the current data do not provide conclusive evidence for SN thermal emission: the possible slight rebrightening could also be attributed to a newly launched jet component \citep{Shu2025}, and the lack of late-time spectroscopy, together with the limited quality of the late-time photometry, means that the data could equally be explained by jet emission alone \citep{Shu2025,Busmann2025}.
Furthermore, SN thermal emission cannot account for the X-ray break, which still requires jet shut-off \citep{Shu2025}.


\subsection{Candidate models and observational constraints}

\begin{table*}[]
\centering
\caption{Possible models for the complex multi-band light curves.}
\begin{tabular}{cccccccc}
\hline
\hline
Model & Ref. & Prompt X-ray & 6-8d rebrighten & 20d rebrighten? & sudden drop & off-nuclear & host SB \\
\hline
refreshed shock            & 1   & SN jet  & refreshed shock     & SN thermal  & ?   & yes & yes \\
structured jet             & 2,3 & SN jet  & jet core appearance & SN thermal  & ?   & yes & yes \\
IMBH rpTDE                 & 4   & 1st TDE & 2nd TDE             & 3rd TDE     & yes & yes & ?   \\
micro-TDE (cluster dynamic) &    & 1st TDE & 2nd TDE             & 3rd TDE     & yes & yes & ? \\
micro-TDE (SN in binary)    &    & SN jet  & TDE of companion    & SN thermal? & yes & yes & yes \\
\hline
\end{tabular}
\begin{tablenotes}
    \item The references are: 1, \cite{Busmann2025}; 2, \cite{Gianfagna2025}; 3, \cite{Quirola-Vasquez2026}; 4, \cite{Shu2025}.
    We list each model's explanations for the prompt X-ray emission, the rebrightening at 6--8 days, and the possible slight rebrightening at $\sim20$ days.
    For constraints from our new observations, a ``yes'' indicates that the model provides a natural explanation, while a ``?'' indicates that the model cannot naturally account for the observation but is not yet ruled out.
\end{tablenotes}
\label{tab:models}
\end{table*}

The most distinctive features of EP241021a are the prominent optical rebrightening at a phase of $\sim7$ days, with an amplitude of $\sim2$ magnitude \citep{Busmann2025}, and the month-long X-ray plateau \citep{Shu2025}.
These two features may share a common origin, as the X-ray light curve also exhibits a rebrightening at a similar time, followed by a power-law decline similar to that in the optical \citep{Shu2025}.
The optical rebrightening is likely of jet origin rather than from SN thermal emission, because the optical spectra taken at phases of 8--9 days exhibits a power-law continuum instead of a blackbody \citep{Shu2025}, and the optical photometry shows persistently red colors \citep{Busmann2025}.

We now list and discuss the models proposed to explain the optical rebrightening and the X-ray plateau.
We also examine whether these models can account for the possible slight optical rebrightening at $\sim20$ days, as well as our new observational results obtained in this work, including the off-nuclear location of the transient and the final sudden drop in both bands.
For the convenience of the reader, we summarize the comparison of these models in Table~\ref{tab:models}.

\subsubsection{SN with a single jet?}

In the standard framework of an SN with a single jet, two models have been proposed to account for the jet-driven optical rebrightening.
In the refreshed shock model, the rebrightening arises from energy injection as a slower trailing shell overtakes and merges with the decelerated forward shock \citep{Busmann2025}.
In the off-axis structured jet model, the rebrightening is produced by the delayed entry of the jet core into the line of sight as the jet decelerates and its beaming cone widens \citep{Gianfagna2025}.

The projected physical offset of $0.95\pm0.13$ kpc is consistent with the spatial distribution of long GRBs, which spans from $\sim0.1$ to several tens kpc with a median of 1.3 kpc \citep{Blanchard2016,WangFF2020}.
The stellar mass and star formation rate of the host galaxy also match those of long GRBs \citep{Quirola-Vasquez2026}.
However, the main difficulty for these models is the sudden drop at $\sim40-50$ days, particularly in the X-ray band (Section 4.1).
Since the central engine of a core-collapse SN lasts only $\lesssim10^3-10^4$ s \citep[e.g.,][]{Woosley2006}, the drop cannot be explained by jet shut-off.
The same problem applies to models invoking a binary compact object merger, whose central engine activity is similarly short-lived \citep[e.g.,][]{Fryer1999}.

\subsubsection{IMBH repeating partial TDEs?}

If EP241021a is a TDE, the accretion can be sustained for far longer than in a core-collapse SN or a binary compact object merger, supporting long-lived jet activity.
In this framework, the rebrightening at $\sim7$ days and the possible slight rebrightening at $\sim20$ days can be interpreted as the second and third jet ejections, and the final drop as jet shut-off.
Such repeated ejections require repeating partial TDEs (rpTDE).
The X-ray luminosity at the break, a few $\times10^{43}$ erg~s$^{-1}$, is lower than those of jetted SMBH TDEs, based on which \cite{Shu2025} constrained the mass of the central engine to $\lesssim10^4$ $M_\odot$ and suggested that it could be an IMBH.
Such an IMBH rpTDE model can account for the complex multi-band evolution, from the prompt emission to the final jet shut-off, as detailed in section 4.4 of \cite{Shu2025}.

The off-nuclear location of EP241021a rules out a nuclear black hole as the central engine, but is consistent with the IMBH scenario, as several off-nuclear IMBH TDE candidates have been reported, including 3XMM J2150-0551 \citep{LinDC2018}, EP240222a \citep{JinCC2025}, and EP250702a \citep{LiDY2026}.
The starburst and the young stellar population ($\lesssim100$ Myr) of the host galaxy are not directly related to the formation of an IMBH, which requires much longer timescales.
Multiwavelength searches have revealed TDEs in starburst galaxies, including obscured infrared TDEs, suggesting an enhanced TDE rate around nuclear SMBHs \citep{Tadhunter2017,Mattila2018}, although whether this enhancement applies to IMBHs on the galactic disk remains uncertain.
Thus, the starburst host neither supports nor rules out the IMBH scenario.

\subsubsection{Micro-TDE?}

In the TDE framework, the central engine could also be a stellar-mass compact object, either a stellar-mass black hole (StMBH) or a neutron star, whose mass is well within the $\lesssim10^4$ $M_\odot$ constraint \citep{Shu2025}.
The tidal disruption of a star by such an object is termed a micro-TDE \citep{Perets2016}.
Simulations of micro-TDEs \citep{Kremer2019,Tsuna2025} show that their peak luminosities can reach $\sim10^{44}$ erg~s$^{-1}$, sufficient to account for the observed plateau luminosities of EP241021a, and that the accretion can last for up to a month, matching the observed jet shut-off timescale.

Two channels have been proposed for micro-TDEs.
In the cluster dynamics channel \citep{Kremer2019,Vynatheya2024,Rastello2026}, dynamical scattering within a dense young cluster delivers a star onto a close orbit around a StMBH, where it is tidally disrupted.
If the orbit is highly eccentric and the pericenter lies only slightly outside the tidal disruption radius, the star can undergo rpTDEs, producing multiple jet ejections.
Such a StMBH rpTDE model, like its IMBH counterpart, can account for all the observed features of EP241021a, with the only difference being the smaller mass of the central engine.

Another channel for a micro-TDE arises in binary systems that produce a Type Ibc supernova \citep{Hirai2022,Tsuna2025}.
An asymmetric explosion imparts a natal kick to the newborn compact object, potentially driving it toward its companion and triggering a TDE.
If the binary is unbound after the kick, or if the companion is fully disrupted, only a single TDE occurs.
If instead the binary remains bound with a pericenter distance in the range that yields a partial TDE, multiple TDE episodes can occur.
In either case, this scenario can explain the complexity in EP241021a's light curve: the prompt X-ray emission originates from the SN jet, while the rebrightening arises from the micro-TDE jet.
The rest-frame time interval of $\sim4$ days corresponds to the travel time of the newborn compact object to its companion, consistent with the theoretical estimates of \cite{Tsuna2025}.
The micro-TDE jet is expected to be oriented according to the binary orbit and is generally misaligned with the SN jet.
This is consistent with the observations: the SN jet may be narrow and pointed toward the observer, accounting for the highly relativistic prompt emission;
whereas the micro-TDE jet may be wider and inclined relative to the line of sight, making it only mildly relativistic with a luminosity of $\sim10^{44}$ erg~s$^{-1}$, close to the thermal luminosity of the micro-TDE.
The possible 20-day slight rebrightening can be understood in two ways: it may be thermal emission from the SN ejecta, or it may be a second TDE episode if rpTDEs occur.

Of the two channels, the SN-in-binary channel is favored by the high star formation rate of the host galaxy.
In contrast, the cluster dynamics channel does not necessarily require a starburst host, since dense stellar clusters can sustain a significant micro-TDE rate over long timescales \citep{Kremer2019}.

Assuming that EP241021a originates from the SN-in-binary channel, we estimate the implied occurrence rate and check whether it is consistent with theoretical expectations.
If most of the 63 X-ray transients detected by EP/WXT in 2024 \citep{ZhangYQ2025} are associated with core-collapse SNe, and if roughly half of massive stars reside in binary systems, then finding one micro-TDE among them implies a micro-TDE fraction of a few percent among binary systems.
\cite{Tsuna2025} estimated a fraction of $\sim0.1\%-1\%$ for relatively wide binary separations ($\sim10^{13}$ cm, corresponding to orbital periods of $\sim10^2$ days).
However, the progenitors of jet-producing SNe Ibc are more likely to reside in tight binaries with orbital periods of $\sim1-10$ days \citep{Fuller2022}.
Closer separations increase the probability of a close encounter or collision, making a micro-TDE fraction of a few percent plausible.

\section{Conclusion} \label{sec:5}

Our new deep late-time imaging data reveal that the host galaxy of EP241021a is a dwarf disk galaxy with $R_e=1.7$ kpc, $M_\star=2.7\pm1.2\times10^8 M_\odot$, and a star formation rate of $4.5\pm2.4M_\odot\ {\rm yr}^{-1}$, consistent with a starburst.
The transient is consistent with residing in the galactic disk, with a projected physical offset of $0.95\pm0.13$ kpc from the galaxy center.
By subtracting the host contribution, we further reveal that the optical rebrightening, previously known to peak at $\sim7$ days, lasted only until $\sim47$ days, followed by a sharp decline ($\alpha_2<-3.3$ at the $3\sigma$ level).
This sharp decline, together with the synchronous drop in the X-ray light curve, points to jet shut-off, which in turn requires a month-long accretion and thus favors TDE interpretations.
We slightly favor the post-SN micro-TDE scenario in a binary system, primarily because it is naturally linked to the starburst.
Nevertheless, we cannot yet exclude rpTDEs by IMBH or StMBH in a dense stellar cluster.

Future EP observations will discover more off-nuclear FXTs with month-long timescales, similar to EP241021a.
These will allow the event rate and host galaxy environments of this intriguing new population to be studied statistically, shedding light on their nature.

\begin{acknowledgments}
We thank the anonymous referee for constructive comments that helped improve the presentation of this paper.
We acknowledge the use of DeepSeek, a large language model developed by DeepSeek AI, for language polishing.
The authors are solely responsible for the content and accuracy of this manuscript.
This research is based on observations made with the NASA/ESA Hubble Space Telescope obtained from the Space Telescope Science Institute, which is operated by the Association of Universities for Research in Astronomy, Inc., under NASA contract NAS 5–26555.
We are grateful to the Director of the Space Telescope Science Institute for awarding Director's Discretionary (DD) time for this program, under program ID 18213.
Based on observations obtained at the International Gemini Observatory, which is operated by the Association of Universities for Research in Astronomy, Inc., under a cooperative agreement with the NSF on behalf of the Gemini partnership: the National Science Foundation (United States), the Science and Technology Facilities Council (United Kingdom), the National Research Council (Canada), CONICYT (Chile), the Australian Research Council (Australia), Minist\'{e}rio da Ci\^{e}ncia e Tecnologia (Brazil), and SECYT (Argentina).
These observations are associated with program GN-2025B-Q-223.

L.-M. Sun is supported by the National Natural Science Foundation of China (NFSC, 12103002).
X.-W. Shu acknowledges the support by National Key R\&D Program of China (No. 2025YFF0511101), 
the National Natural Science Foundation of China through grant No. 12192221, and 
the science research grants from the China Manned Space Project with CMS-CSST-2025-A07.
J.-H. Chen acknowledges the National Natural Science Foundation of China (the NSFC Type C Youth Project 12503053).
C.-C. Jin acknowledges the National Natural Science Foundation of China through grant 12473016, the National Key R\&D Program of China No. 2025YFF0511100, and the Strategic Priority Research Program of the Chinese Academy of Sciences (Grant No. XDB0550200).
\end{acknowledgments}


\facilities{HST(WFC3), Gemini-N(GMOS)}

All the {\it HST}/WFC3 data used in this paper can be found in MAST: 
\dataset[10.17909/965f-5702]{https://doi.org/10.17909/965f-5702}.

\software{
astropy \citep{2022ApJ...935..167A},
emcee \citep{2013PASP..125..306F},
matplotlib \citep{2007CSE.....9...90H},
CIGALE\citep{Boquien2019},
SExtractor\citep{Bertin1996},
Scamp\citep{Bertin2006}
          }


\bibliography{EP241021a_HST}
\bibliographystyle{aasjournalv7}

\end{document}